\documentclass[10pt,twocolumn,twoside]{elsart}

\usepackage{graphicx}
\usepackage{amsmath}

\begin{document}

\begin{frontmatter}

\title{Di-hadron azimuthal correlation and Mach-like cone structure in a parton/hadron transport model}

\author{G. L. Ma$^{a,b}$, S. Zhang$^{a,b}$, } \author{  Y. G.
Ma$^{a}$} \ead{ygma@sinap.ac.cn. Corresponding author },
\author{H. Z. Huang$^c$, X. Z. Cai$^{a}$, }
\author{ J. H. Chen$^{a,b}$, Z. J. He$^a$, J. L. Long$^a$, W. Q. Shen$^{a}$, X. H. Shi$^{a,b}$, J. X. Zuo$^{a,b}$ }
\author{ }

\address{$^a$ Shanghai Institute of Applied Physics, Chinese Academy of
Sciences, Shanghai 201800, China\\
$^b$ Graduate School of the Chinese Academy of Sciences, Beijing
100080, China\\ $^c$ University of California, Los
Angeles,CA90095, USA}

\date{\today}

\begin{abstract}

In the framework of a multi-phase transport model (AMPT) with both partonic and hadronic interactions, azimuthal
correlations between trigger particles and associated scattering particles have been studied by the mixing-event
technique. The momentum ranges of these particles are $3< p^{trig}_T< 6$ GeV/$c$ and $0.15< p_{T}^{assoc} < 3$
GeV/$c$ (soft), or $2.5<p^{trig}_T<$ 4 GeV/$c$ and $1< p_{T}^{assoc} < 2.5$ GeV/$c$ (hard) in Au + Au collisions
at $\sqrt{s_{NN}}$ = 200 GeV. A Mach-like structure has been observed in correlation functions for central
collisions. By comparing scenarios with and without parton cascade and hadronic rescattering, we show that both
partonic and hadronic dynamical mechanisms contribute to the Mach-like structure of the associated particle
azimuthal correlations. The contribution of hadronic dynamical process can not be ignored in the emergence of
Mach-like correlations of the soft scattered associated hadrons. However, hadronic rescattering alone cannot
reproduce experimental amplitude of Mach-like cone on away-side, and the parton cascade process is essential to
describe experimental amplitude of Mach-like cone on away-side. In addition, both the associated multiplicity
and the sum of $p_{T}$ decrease, whileas the $\langle p_{T} \rangle$ increases, with the impact parameter in the
AMPT model including partonic dynamics from string melting scenario. \vspace{1pc}
\end{abstract}


\end{frontmatter}

\begin{description}
   \item[I. Introduction ]
\end{description}

A phase transition between hadronic matter and the quark-gluon
plasma (QGP) at a critical energy density of $\sim$ 1 $GeV/fm^{3}$
has been predicted by Quantum Chromodynamics (QCD)~\cite{QCD},
which has motivated the scientific program at the Relativistic
Heavy-Ion Collider (RHIC) at the Brookhaven National Laboratory. A
very dense partonic matter has been shown to be produced in the
early stage of central Au+Au collisions at $\sqrt{s_{NN}}$ = 200
GeV~\cite{White-papers}. Many interesting phenomena have been
observed including the measurements of elliptic
flow~\cite{ellipticflow}, strangeness~\cite{STARphi},
J/$\psi$~\cite{jpsi} and jet quenching~\cite{jet-ex}.

Jet has been proved as a particularly good probe in RHIC experiments~\cite{jet-ex}.
The high transverse momentum ($p_{T}$) partons (jets)
which emerge from hard scattering processes will lose energy when
they pass through the dense QCD medium. The energy loss (jet
quenching) mechanism results in distinct experimental observations
such as the disappearance of one jet in back-to-back jet
correlation at high $p_{T}$~\cite{hard-hard-ex}. At the same time,
the loss energy must be redistributed in the soft $p_T$
region~\cite{soft-soft-th1,soft-soft-th2,soft-soft-th3,soft-soft-th4}.
Experimentally the soft scattered particles which carry the lost
energy have been observed statistically via two-particle angular
correlation of charged particles~\cite{soft-soft-ex}.
Reconstruction of these particles will constrain models which
describe production mechanisms of high $p_{T}$ particles, and may
shed light on the underlying energy loss mechanisms and the degree
of equilibration of jet products from energy loss in the medium.

A Mach-like structure (the splitting of the away side peak in di-jet correlation) has recently been observed in
azimuthal correlations of scattered secondaries associated with the high $p_{T}$ hadrons in central Au+Au
collisions at $\sqrt{s_{NN}}$ = 200 GeV~\cite{sideward-peak1,sideward-peak2,sideward-peak3}. Several theoretical
interpretations have been proposed for this phenomenon. For instances, St\"ocker  proposed Mach cone structure
from jets traversing the dense medium as a probe of the equation of state and the speed of sound of the
medium~\cite{Stocker}. Casalderrey-Solana, Shuryak and Teaney argued for a shock wave generation because jets
travel faster than the sound in the medium~\cite{Casalderrey}. They fit the broad structure on the away side of
the azimuthal correlation with a Mach cone (shock wave) mechanism; Koch and Wang et al. could produce a
Mach-like structure with a Cherenkov radiation model~\cite{Koch}; Armesto et al. interpreted the sideward peaks
as a result of medium dragging effect in Ref.~\cite{Armesto}; Ruppert and M\"uller argued that the Mach-like
structure can appear due to the excitation of collective plasmon waves by the moving color charge associated
with the leading jet~\cite{Ruppert}; Renk and Ruppert applied a realistic model for the medium evolution to
explain the observed splitting of the away side peak in Ref.~\cite{Thorsten}; Chaudhuri studied the effect on
Mach-like structure from jet quenching on the hydrodynamical evolution of a QGP fluid~\cite{Chaudhuri}; Satarov,
St\"ocker and Mishustin investigated Mach shocks induced by partonic jets in expanding quark-gluon
plasma~\cite{Sat}; A conical flow induced by heavy-quark jets was also proposed by Antinori and Shuryak
\cite{Anto}; Hwa et al. discussed the centrality dependence of associated particle distribution for Au + Au and
d + Au, respectively, in recombination model~\cite{Hwa}. The dynamical nature of the Mach-like structure
continues to be a subject of many theoretical and experimental investigations.

There is no quantitative interpretation from dynamical transport models yet
for the Mach-like structure in particle correlations
from hard scattering particles interacting with the dense medium.
In this work, we report a study of
associated particle correlations with a triggered particle and investigate the Mach-like
structure using A Multi-Phase Transport
model( AMPT )~\cite{AMPT}. We have applied the mixing-event technique in the
AMPT analysis as what has been used in the analysis of RHIC collision data to remove background.
We found that both parton cascade
and hadronic rescattering can produce apparent correlations between triggered and associated
particles similar to the Mach-like structure. But the
hadronic rescattering mechanism alone is not able to produce a large
enough amplitude for Mach-like cone on away side, and the parton
cascade process is indispensable.

The paper is organized as follows. In Section II, we give a brief
description of the AMPT model and the initial conditions. Section
III describes  mixing-event technique in our simulation analysis.
Results and discussions are presented in Section IV.  Finally
a summary is given in Section V.

\begin{description}
   \item[II. Brief description of AMPT Model ]
\end{description}

The AMPT model~\cite{AMPT} is a hybrid model which consists of four main components: the initial conditions,
partonic interactions, the conversion from partonic matter into hadronic matter and hadronic rescattering
interactions. The initial conditions, which include the spatial and momentum distributions of minijet partons
and soft string excitation, are obtained from the HIJING model~\cite{HIJING}. Excitation of strings will melt
the string into partons. Scatterings among partons are modelled by Zhang's parton cascade model
(ZPC)~\cite{ZPC}, which includes two-body scatterings with cross sections obtained from the pQCD calculations
with screening mass. In the default AMPT model~\cite{DAMPT} partons are recombined with their parent strings
when they stop interacting, and the resulting strings are converted to hadrons by using the Lund string
fragmentation model~\cite{Lund}. In the AMPT model with the option of string melting~\cite{SAMPT}, a quark
coalescence model is used to combine partons into hadrons. Dynamics of the subsequent hadronic matter is then
described by A Relativistic Transport (ART) model~\cite{ART}. Details of the AMPT model can be found in a recent
review~\cite{AMPT}. Previous studies~\cite{SAMPT} have shown that the partonic effect could not be neglected and
the string melting AMPT is much more appropriate than the default AMPT version when the energy density is much
higher than the critical density for the predicted phase transition~\cite{AMPT,SAMPT,Jinhui}. In the present
work, the parton interaction cross section in AMPT model with string melting is assumed to be 10mb.

\begin{description}
   \item[III. Analysis Method  ]
\end{description}

\begin{figure}
\includegraphics[scale=0.55]{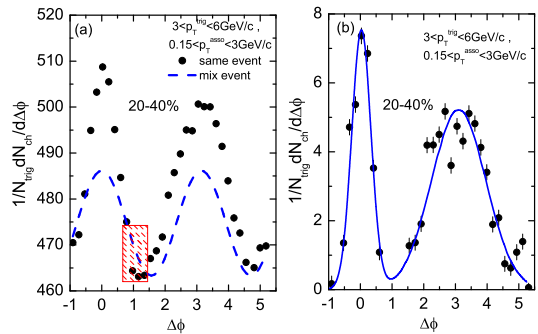}
\caption{\footnotesize (a): The $\Delta\phi$ distribution  between a triggered hadron ($3<p_{T}^{trig}<6$
GeV/$c$) and associated hadrons ($0.15<p_{T}^{assoc}<3$ GeV/$c$) (circles), where the background (dash line) has
not been subtracted. The data are 200 GeV Au + Au collisions with 20-40\% collision centrality from the AMPT
model. The dash area is the region of ZYAM normalization (see texts for detail); (b): The $\Delta\phi$
distribution where the background has been subtracted by a mixing-event technique, where the solid line is a
two-Gaussian fit.
 }
\label{method}
\end{figure}

In order to compare with experimental measurements of correlations
between trigger and associated hadrons, we use the mixing-event
technique to subtract combinatorial background in our analysis.
Two ranges of  $p_{T}$ window selections for the trigger and
associated particles have been used: one is $3 < p_{T}^{trig} < 6$
GeV/$c$ and $0.15< p_{T}^{assoc} < 3$ GeV/$c$ (we call this
selection as "soft" associated hadrons since soft particles
dominate in this $p_T$ range for associated particles); the other
is $2.5 < p_{T}^{trig} < 4$ GeV/$c$ and $1.0 < p_{T}^{assoc} <
2.5$ GeV/$c$ (denoted as "hard" associated hadrons since there are
more hard particle components than that in the "soft" selection).
The triggered particles and associated particles ("soft" and
"hard" ones) both are selected with a pseudo-rapidity window
$|\eta| < 1.0$. In the same events, pairs of associated particles
with a triggered particle are accumulated to obtain $\Delta\phi =
\phi - \phi_{trig}$ distributions. In order to construct the
background which is mainly from the effect of elliptic flow
\cite{soft-soft-ex,sideward-peak2}, a mixing-event method is
applied to simulate the background. In this method, we mixed two
events which have very close centrality into a new event, and
extracted $\Delta\phi$ distribution to be used as background
distribution. When subtracting the background from the same
events, ZYAM (zero yield at minimum) assumption is adopted which
has been used in the experimental analysis~\cite{sideward-peak2}.
Figure~\ref{method} shows the $\Delta\phi$ distribution for
trigger hadrons of $3 < p_{T}^{trig} < 6$ GeV/$c$ and associated
hadrons of $0.15 < p_{T}^{assoc} < 3$ GeV/$c$  before and after
the background subtraction. The data are from AMPT model
simulation of Au + Au collisions at $\sqrt{s_{NN}}$ = 200 GeV for
20-40\% collision centrality, where the string melting mechanism
has been turned on.

\begin{description}
   \item[IV. Results and Discussions]
\end{description}

\begin{figure}
\includegraphics[scale=0.50]{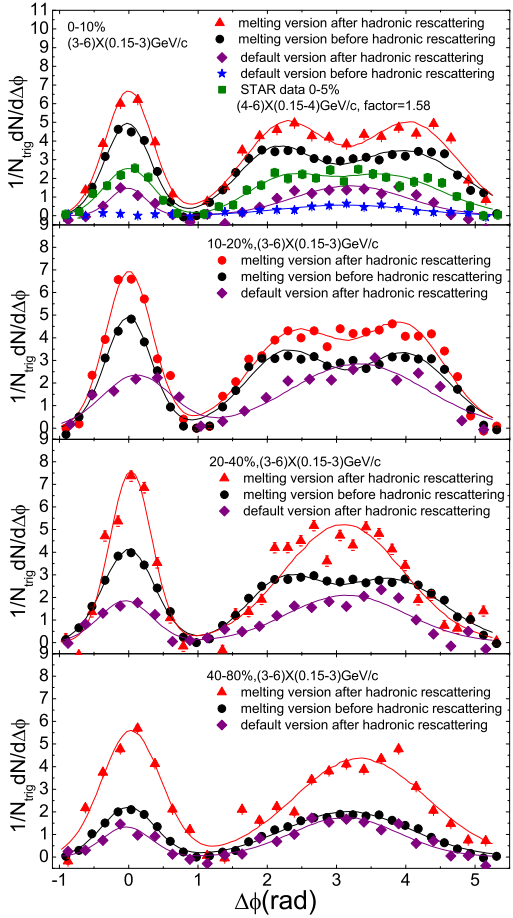}
\caption{\footnotesize Soft scattered associated hadron
$\Delta\phi$ correlations for trigger hadrons of $3.0 <
p_{T}^{trig} < 6.0 $GeV/$c$  and associated hadrons of $0.15 <
p^{assoc}_T < 3.0 $ GeV/$c$ from AMPT Monte Carlo simulations of
Au + Au collisions at $\sqrt{s_{NN}}$ = 200 GeV  with various collision centralities.
Triangles: string melting version after hadronic rescattering;
circles: string melting version before hadronic rescattering; diamonds:
default version after hadronic rescattering; stars: default
version before hadronic rescattering; squares: experimental data are
from Ref~\cite{soft-soft-ex} where $4.0 < p^{trig}_{T} < 6.0 $
GeV/$c$ and $0.15 < p^{assoc}_T < 4.0$ GeV/$c$.
 }
\label{softCor_machcones}
\end{figure}

In Ref.~\cite{soft-soft-ex}, it was found that the $\Delta\phi = \phi - \phi_{trig}$
distribution of recoiling hadrons from a high $p_{T}$ triggered particle
is significantly broadened in
central Au + Au collisions at $\sqrt{s_{NN}}$ = 200 GeV, which supports
the picture of the dissipation of jet energy in the medium. In order to
increase the statistical sample of triggered particles in our
calculation, we set $p_{T}$ range for trigger particles to $3 <
p_{T}^{trig} < 6$ GeV/$c$ and for associated particles to $0.15 <
p_{T}^{assoc} < 3$ GeV/$c$ in our analysis. Both triggered and
associated particles are further selected with a pseudo-rapidity cut of
$|\eta| < 1.0$.

Figure~\ref{softCor_machcones} presents AMPT model calculations of soft
associated hadron $\Delta\phi$ correlations from Au + Au collisions
at $\sqrt{s_{NN}}$ = 200 GeV for different centralities under
various AMPT running conditions. In order to compare our results with
experimental data which give the correlations among associated
charged hadrons, the experimental data are multiplied by a factor
of 1.58 to account for the contribution from neutral hadrons
\cite{soft-soft-ex,factor}. Note that for the default AMPT version
before hadronic rescattering, we here give $\Delta\phi$
correlation for 0-10\% collision centrality only due to the lack of statistics
obtained for other centrality bins. We found
that the hadronic rescattering increases  $\Delta\phi$ correlation
yields for both versions.  A very strong Mach-like structure is
observed for central Au+Au collisions before hadronic rescattering
in the melting AMPT version, which indicates that the Mach-like
structure has been formed in parton cascade process. The effect of
hadronic rescattering on the $\Delta\phi$ correlation in central
collisions ($<20\%$ centrality) does not wash out, even slightly enhances, the Mach-like
structure, which is
qualitatively in agreement with the effect of time-dependent speed of
sound on the development of the conical wave in expanding QCD
matter~\cite{variblespeed}.

\begin{figure}
\includegraphics[scale=0.50]{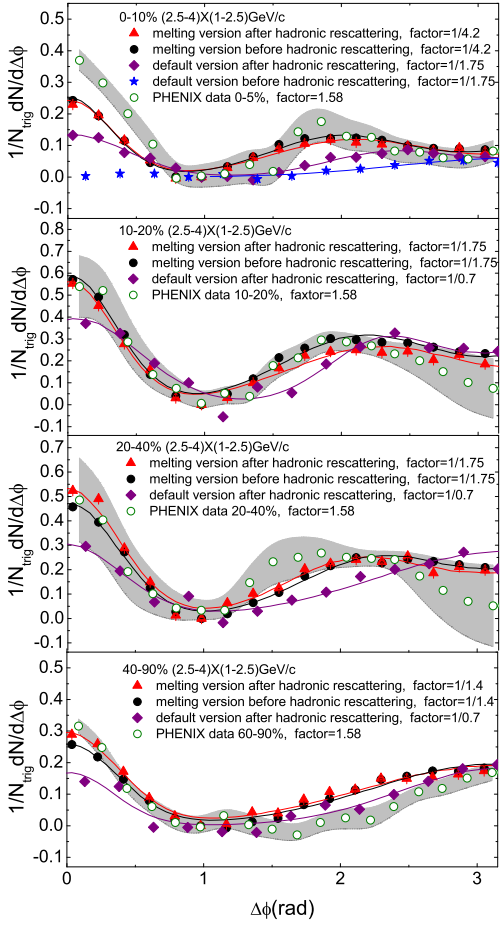}
\caption{\footnotesize AMPT model calculations of $\Delta\phi$ correlations between a triggered hadron of $2.5 <
p^{trig}_T < 4.0$ GeV/$c$ and hard associated hadrons of $1.0 < p_{T}^{assoc} < 2.5$ GeV/$c$ from Au + Au
collisions at $\sqrt{s_{NN}}$ = 200 GeV in various centralities. Triangles: AMPT version with string melting
after hadronic rescattering; Full circles: AMPT version with string melting before hadronic rescattering;
Diamonds: default AMPT version after hadronic rescattering; Stars: default AMPT version before hadronic
rescattering; Open circles: experimental data from Ref.~\cite{sideward-peak2}; Hatched areas indicate the
experimental uncertainty. Note that the scaling factors may also reflect different pseudorapidity ranges used in
experimental data analysis and model calculations. See text for details.
 }
\label{hardCor_machcones}
\end{figure}

Figure~\ref{hardCor_machcones} shows the AMPT calculations of $\Delta\phi$ correlations between a triggered
particle of $2.5 < p^{trig}_T < 4.0 $GeV/$c$ and hard associated particles of $1.0 < p_{T}^{assoc} < 2.5$
GeV/$c$ from Au + Au collisions at $\sqrt{s_{NN}}$ = 200 GeV for various centralities, where pseudorapidity cuts
of $|\eta^{trig}| < 1.0$ and $|\eta^{assoc}| < 1.0$) have also been applied. We found that the effect on
$\Delta\phi$ correlations from hadronic rescattering is much smaller than the case for soft associated
particles, which may indicate that fewer fraction of hard associated hadrons suffer hadronic rescattering. In
addition, Mach-like structures have been observed on away-side correlations in both string-melting and default
AMPT versions. In the default AMPT version, however, the Mach-like structures can be observed only after turning
on hadronic rescatterings.

The default AMPT version appears to produce the number
of associated particles matching better with the experimental measurement
than that from the string-melting AMPT version. This may be
attributed to the fact that the string-melting AMPT model always produces softer $p_{T}$
spectra than the default AMPT model because current quark
masses have been used in the partonic cascade stage~\cite{AMPT}. It may
be improved if thermal parton masses are applied~\cite{Ko-private}.
However, in the present study, we only focus on the correlation shape instead of yields in the
correlation region.  In this context, we shall compare the
shapes of $\Delta\phi$ correlation functions between the AMPT model and
experimental data. To this end, the AMPT model data have been multiplied
with different normalization factors which are listed on right sides
of each panel in Figure~\ref{hardCor_machcones}. Please note that there
is a difference for pseudo-rapidity cut between our simulation and
the experimental data. Our simulation cut of associated particle is $|\eta| < $
 1.0 (due to the limited statistics),while $|\eta| < $ 0.35 in PHENIX data.
In this case, the scaling factors of our simulation should be
reduced by a factor of about 3 in order to match the data. By this
reduction, the scaling parameters shown in the figures have been
normalized to the same $|\eta|$ cuts as the data. However, even with the scaling of 3
these factors still deviate from 1 and show a centrality
dependence. These deviations can be attributed to the following facts:
(1) a large yield of particles in the correlation region can partially come from the
fact that our $p_T$ spectra in AMPT calculation with string
melting scenario is under-predicted in high $p_T$ region if we
compare with the experimental data. In this case, the soft component
will be over-predicted if we compare $\frac{dN}{N_{trig}d\Delta
\phi}$ between the data and simulation; (2) the centrality
dependence of scaling factors may stem from more excessive parton
interactions when a jet parton passes through various dense
partonic matter depending on the collision centrality.
In this study, our emphasis will be on the comparison of shapes of
the Mach-like structure between the AMTP model calculations and experimental data.
Our simulation results indicate that the string-melting AMPT model can describe shapes of
$\Delta\phi$ correlations between a triggered hadron and associated particles
better than the default AMPT model, especially for central Au+Au collisions.

\begin{figure}
\includegraphics[scale=0.40]{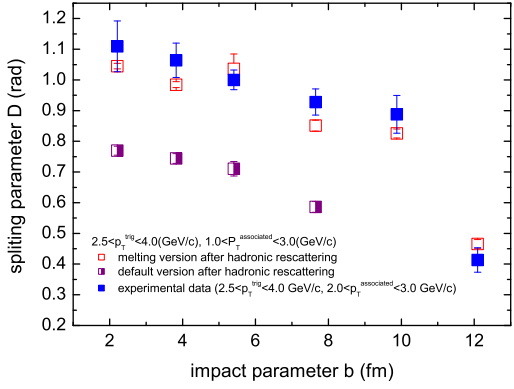}
\caption{\footnotesize Splitting parameter $D$ versus impact
parameter for Au + Au collisions at $\sqrt{s_{NN}}$ = 200 GeV
($p_T$ windows in model: $2.5 <p_{T}^{trig}< 4.0$ GeV/$c$ and $1.0
<p_{T}^{assoc}< 3$ GeV/$c$; $p_T$ windows in experiments: $2.5
<p_{T}^{trig}< 4$ GeV/$c$ and $2.0 <p_{T}^{assoc}< 3.0$ GeV/$c$).
Open symbols: the string-melting version with  hadron rescatterings; half
filled symbols: the default version with hadronic rescatterings;
filled symbols: experimental data from Ref.~\cite{sideward-peak3}.
 }
\label{Danalysis}
\end{figure}

In order to quantitatively characterize the Mach-like structure, a splitting parameter $D$ has been  extracted
in our analysis. The splitting parameter $D$ is defined as half distance between two Gaussian peaks on away side
of associated particle $\Delta\phi$ correlations. The $D$ value reflects the size of direction-splitting of
Mach-cone on the away side.  Figure~\ref{Danalysis} shows the impact parameter dependence of $D$. Our results
indicate that the string-melting AMPT version can roughly match the experimental data. Note that of our $p_T$
cuts are slightly different from experimental cuts in order to increase our statistics. However, $D$ values from
the default AMPT version are significantly smaller than the experimental data. We conclude that hadronic
rescattering mechanism alone is not enough to produce the amplitude of Mach-like cone structure on the away side
and parton cascade mechanism is necessary. It should be noted that minijet partons produced from hard
scatterings can lose energy by gluon radiations and transfer their energies to nearby soft strings in the HIJING
model, which has been called jet quenching \cite{HIJING}. In the AMPT model, the jet quenching in the HIJING
model is replaced by parton scatterings from the ZPC. Only two-body scatterings are included in the current ZPC,
and higher-order contributions such as 2$\rightleftharpoons$3 scattering processes to the jet energy loss are
still missing in our AMPT calculations. It has been proposed that higher order processes may contribute
significantly to the observed large jet quenching~\cite{zhexu}. However our results indicate that
2$\rightleftharpoons$2 processes could account for most of the amplitude for the Mach-like structure. The AMPT
version with string melting scenario has also been shown to generate elliptic flow of hadrons better than the
default version. Furthermore, it can reproduce the mass ordering in the particle dependence of elliptic flow,
which can be described by hydrodynamics models~\cite{SAMPT,Jinhui}. It is believed that the big cross section
for parton interactions used in the AMPT model leads to strong parton cascades that couple partons together,
resulting the onset of hydrodynamics behavior from parton cascades~\cite{Zhang99}. It has also been proposed
that a Mach-like structure can be generated if the jet velocity is faster than the sound velocity when the jet
traverses through the dense matter~\cite{Stocker,Casalderrey}. In addition,  continuous partonic rescatterings
and the rapid expansion of the dense medium can also lead to parton's direction in the medium to both sides of
triggered particle direction, which may result in an apparent Mach-like structure in the AMPT model.
Nevertheless, we conclude that significant parton cascades are essential in order to describe elliptic flow of
hadrons and the Mach-like structure in particle $\Delta\phi$ correlations simultaneously.

\begin{figure}
\includegraphics[scale=0.38]{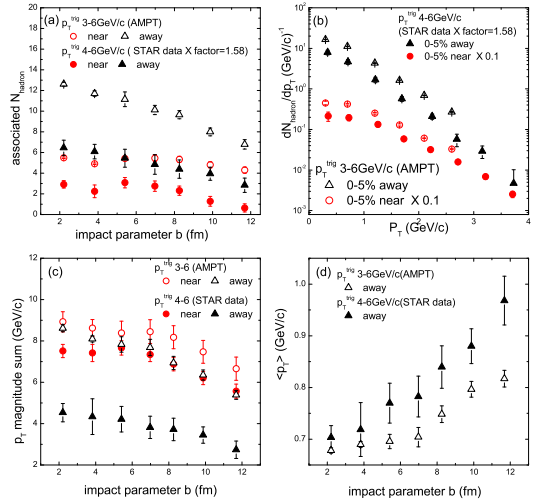}
\caption{\footnotesize Comparisons between model calculations
($p_T$ windows: $3<p_{T}^{trig}<6$ GeV/$c$ and
$0.15<p_{T}^{assoc}<3$ GeV/$c$) and experimental data
\cite{soft-soft-ex} ($p_T$ windows: $4 <p_{T}^{trig}< 6$ GeV/$c$
and $0.15 <p_{T}^{assoc}< 4$ GeV/$c$) in Au + Au collisions at
$\sqrt{s_{NN}}$ = 200 GeV. Open symbols: calculations from the
string-melting AMPT version, filled symbols: experimental data, circle:
near side, triangle: away side. Panel (a): Number of associated
hadrons versus impact parameter; Panel (b): $p_{T}$ distributions
for the associated hadrons  on near side and away side for 0-5\%
centrality; Panel (c): Sum of $p_{T}$ magnitude as a function of
impact parameter on near side and away side; Panel (d): $\langle
p_{T}\rangle$ as a function of impact parameter on away side.
 }
\label{Nch-soft}
\end{figure}

We have also investigated several characteristics of soft
associated particles in the AMPT model with the string-melting
scenario. Figure ~\ref{Nch-soft} shows a comparison between
experimental data and our AMPT calculation results with
hadronic rescatterings.
Panel (a) in Fig.~\ref{Nch-soft}
shows the number of associated hadrons as a function of impact
parameter in Au+Au collisions at $\sqrt{s_{NN}}$ = 200 GeV. The
trigger particles are not included in the $N_{ch}$ counting for the near side.
The number of average associated hadrons decreases with increasing impact
parameters in the model as well as in the experiment. However, the
number of associated hadrons in the AMPT model is much larger
than the experimental data on both near and away sides, which may
due to different $p_T$ cuts used in the model calculation and experimental data.
Panel (b) shows the $p_{T}$ distributions of the
associated hadrons on near side and away side in most central
(0-5\% centrality) Au + Au collisions at $\sqrt{s_{NN}}$ = 200
GeV. Panel (c) gives the sum of $p_{T}$ magnitude, which
approximates to associated energy, as a function of impact
parameter. The triggered particles are included in the $p_{T}$ magnitude
sum for the near side. The sums of $p_{T}$ magnitude
decrease with impact parameters for both near side and away side,
and theoretical values are somewhat higher than the experimental
data, again may be affected by the different $p_T$ cuts.
Panel (d) displays the dependence of $\langle p_{T}\rangle$
on impact parameters on away side. $\langle p_{T}\rangle$ increases
with impact parameter.
These comparisons may help to address the issue of parton thermalization through the parton
cascade mechanism in central Au+Au collisions~\cite{soft-soft-ex}.

\begin{description}
   \item[V. Conclusions ]
\end{description}

In summary, the origin of the Mach-like structure in correlations
between triggered hadrons and soft or
hard associated particles has been investigated in the
framework of a hybrid dynamics transport model which includes two
dynamical processes, namely parton cascade and hadronic
rescattering.  By comparing the different calculation results with
or without parton cascade, before or after hadronic rescattering,
we found that the associated particle correlations and the
Mach-like structure have been formed mostly before hadronic rescatterings,
which indicates that these kinds of correlations  are born in the
partonic process and are further developed in the later hadronic
rescattering processes. For the hard associated particles,
hadronic rescattering hardly changes the Mach-like structure which is
mostly formed in parton cascade processes. Therefore, characteristics of hard
associated particles may directly reflect information of
intrinsic partonic dynamics. Meanwhile, the effect of hadronic
rescattering can not be ignored especially for soft
associated particles, because many of these soft associated
particles either are produced in or suffer from hadronic rescattering processes.
Our calculations indicate that hadronic rescattering mechanism alone is
unable to produce a splitting parameters $D$ distribution for the Mach-like
structure matching the experimental data. The parton cascade mechanism
is essential for the Mach-like structure while the exact shape
and strength of the azimuthal correlations seem to depend on detailed
properties of the partonic medium and jet, which still awaits quantitative explanations.

\begin{description}
   \item[Acknowledgements ]
\end{description}We thank Prof. Che-Ming Ko for careful reading of the
manuscript and for many useful suggestions. This work was
supported in part by the Shanghai Development Foundation for
Science and Technology under Grant No. 05XD14021, 06JC14082,  and
03QA14066, the National Natural Science Foundation of China under
Grant No. 10535010, 10328259 and 10135030.


\end{document}